\documentstyle[aps,manuscript]{revtex}

\begin{document}

\title{Theory of the evolutionary minority game}
\author{T.S. Lo$^{1}$, P.M. Hui$^{1}$, and N.F. Johnson$^{2}$}
\address{$^{1}$ Department of Physics, The Chinese University of Hong
Kong,\\
Shatin, New Territories, Hong Kong.\\
$^{2}$ Department of Physics, University of Oxford, \\ 
Clarendon Laboratory, Oxford OX1 3PU, England, UK.}

\maketitle

\begin{abstract}
We present a theory which describes a recently introduced model of an
evolving, adaptive system in which agents compete to be in the minority. The
agents themselves are able to evolve their strategies over time in an
attempt to improve their
performance. The present theory explicitly demonstrates the
{self-interaction}, or {\em market impact}
that agents in such systems experience.


\end{abstract}

\newpage
\section{Introduction}

Agent-based models of complex adaptive systems (CAS) provide invaluable
insight
into the highly non-trivial global behaviour of a population of  competing
agents\cite{holland}.  These models typically involve agents  with
similar capability competing for a limited resource.  The agents  are given
the
same global information, which is in turn generated by  the action of the
agents
themselves, and they learn from past  experience.  The growing field of
econophysics\cite{arthur,stanley,confs} represents an area in which such CAS
may be
applicable: every agent knows the past ups and downs in the
index of a stock market and must decide how to trade based on this global
information. An important step forward in agent-based models of CAS was made
by
Challet and Zhang\cite{challet,savit} who proposed the so-called Minority
Game
(MG) in which an odd number $N$ of agents  successively compete to be in the
minority. Each  agent is randomly assigned
a limited number of strategies at the beginning of the game, hence introducing
some quenched disorder. As the game progresses, non-trivial fluctuations
arise in
the collective agents' decisions -- these can be
understood in terms of  the dynamical formation of crowds consisting of
agents using correlated strategies, and anticrowds consisting  of agents
using
the anticorrelated strategies\cite{crowd}.
Subsequent work by Challet and co-workers
has provided a remarkable formal connection to spin
glass systems\cite{spinglass}.

The basic minority game, however, does not incorporate evolution. Agents are
stuck with their initial strategies and hence the system cannot avoid this
in-built frustration. In the real world, one would expect that agents
would be able to evolve more successful strategies, or at least stop
playing disasterous strategies. This motivated us to recently propose
a simpler minority model which allowed for an {\em evolving} population
\cite{prl,royal,HLJ} - we call this the evolutionary minority game (EMG).
D'Hulst
and Rodgers\cite{dHR} subsequently proposed an analytic theory, based on a
slightly modified version of our model.  However, the two models actually
give
different numerical results \cite{HLJ}.

Here we provide a theory for
our evolutionary minority game (EMG) \cite{prl} which correctly includes the
self-interaction of the agents. Results are in
good
agreement with numerical data.  The plan of the paper is as follows.   We
introduce the EMG in Sec.II and give the main features observed in
numerical
simulations of the model.  In Sec.III,  we present the formalism and derive
the
winning probability for an agent. Results from the present theory are
compared with
numerical data in Sec.IV.  Section V provides a discussion of the results.

\section{Evolutionary minority game}

Consider an odd number
$N$ of agents repeatedly choosing to be in room 0 (e.g. sell) or room 1
(e.g. buy). After each agent has independently chosen a room,
the winners are those
in the minority room.  A single  binary digit denoting the minority room
forms the
outcome for each time-step.   Each agent is given the information of the
most
recent $m$ outcomes.  Each agent also has  access to a common register or
``memory" containing the outcomes  from the most recent occurrences of all
$2^{m}$ possible bit strings  of length $m$. Consider, for example, $m=3$
and
denote ($xyz$)$w$ as the $m=3$ bit string ($xyz$)  and outcome $w$.  An
example
memory would comprise (000)1, (001)0, (010)0, (011)1, (100)0, (101)1,
(110)0,
(111)1.  Following a run of three wins for room 0 in the recent past, the
winning room was subsequently 1.   Faced with a given bit string of length
$m$, it seems reasonable for an agent to simply predict the same outcome as
that
registered in the memory. The agent will hence choose room 1 following the
next 000 sequence.  If 0 turns out to be the winning room, the entry (000)1
in
the memory is then updated to be (000)0.  Simply put, each agent looks into
the
most recent history for the same pattern of $m$ bit string and predicts the
outcome using the history.  In effect, each agent holds one strategy  and
all
agents hold the same strategy, with the strategy being {\em dynamical}.
The
strategy is hence to follow the trend. However, if all $N$ agents act in the
same way, they will all lose.  A successful agent is one who  can follow a
trend
as long as it is valid and to correctly predict when it will end.  To
incorporate
this factor into our model, each agent is assigned a single number $p$,
which we
refer to as the ``gene"-value.  Following a given $m$-bit sequence, $p$ is
the
probability that the agent will choose the same outcome as that stored  in
the
memory, i.e. he will follow the current predictor.  An  agent will reject
the
prediction and choose the opposite  action with probability $1-p$.
To incorporate evolution into our model, we assign $+1$ ($-1$) point to
every
agent in the minority (majority) room at each time step.  If an  agent's
score
falls below a value $d$ ($d<0$), his gene-value $p$  is modified.  The new
$p$
value is chosen randomly from a range of values  centered on the old $p$
with a
width equal to $R$.  We impose  reflective boundary condition to ensure that
$0\leq p\leq 1$.  Our  conclusions do not depend on the particular choice of
boundary  conditions.  For $R=0$, the agents will never change their gene
values - this represents the limiting case of in-built
quenched disorder determined by the initial
distribution of
$p$ values.  For any non-zero $R$ value, the system is able to evolve
through gene
modification. For
$R=2$, the new gene value is uncorrelated with the old one upon
modification.

Initially, each agent is randomly assigned a  gene value in the range $0
\leq p
\leq 1$. Choosing $R\neq 0$ allows the population to evolve.  We focus on
two
quantities,
$P(p)$ and $L(p)$,  in the asymptotic limit.  Here $P(p)$ is the frequency
distribution  of gene values, typically taken in the long time limit over a
time
window and normalized to unity; $L(p)$ is the lifespan defined as  the
average
length of time a gene  value $p$ survives between modifications. To
introduce the
basic features observed in numerical simulations, Fig. 1 shows $L(p)$ and
$P(p)$
(inset) as a function of
$p$ for a range of values of $m$.   The other parameters are taken to be
$N=101$,
$R=0.2$ and $d=-4$. The most interesting feature is that $P(p)$ becomes
peaked
around $p=0$ and
$p=1$, with a similar behaviour in $L(p)$.  Both of these quantities
are symmetric about
$p=1/2$.  The results are insensitive to the initial distribution of $p$
values.
Surprisingly the results indicate  that agents who either always follow or
never
follow what happened last  time, generally perform better than cautious
agents
using  an intermediate value of $p$.  Figure 1 also shows that there is no
explicit dependence on $m$ for $P(p)$ and $L(p)$\cite{prl,HLJ,ceva1}.  The
independence on $m$ of the results was also discussed recently by Burgos and
Ceva\cite{ceva1} using a random walk argument.   Reference \cite{dHR}
proposes a
theory which gives a $P(p)$ somewhat similar to that shown in Fig.1.
However, the theory was developed based on a model in
which each agent is initially assigned one strategy from the strategy pool,
and uses this strategy throughout the game: the corresponding $P(p)$ is then
$m$-dependent\cite{HLJ} in contrast to the EMG results shown in Fig.1.
The dependences
on the other parameters of the EMG such as $N$, $d$, and $R$ are reported in
Ref.\cite{HLJ}.

\section{Formalism}

We consider a game with $N$ agents ($N \gg 1$).  After a sufficiently  long
time,
the distributions $P(p)$ and $L(p)$ reach the stationary  forms as shown in
Fig.
1.  Consider a certain moment of the game in  this steady-state regime.  Let
the
predictor, which is simply
the strategy stored in the memory for the given history bit-string, be
$1$; i.e. go to room ``1". As long as the  winning room is defined
as the minority room, i.e. with a cutoff at $(N-1)/2$, the following
arguments
do not depend on the actual value  of the predictor and hence also hold if
the
predictor says
$0$.  We define $F_{N}(n)$ as the probability of the attendance being $n$ in
the predicted
room.
It follows from the
central limit theorem that
$F_{N}(n)$ will be an approximately gaussian distribution
with a mean $N\overline{p}$ and variance $N \int_{0}^{1} P(p) p (1-p) dp$.  
Here
$\overline{p}$ is the mean of the gene value $p$ given by
$\overline{p} = \int_{0}^{1} p P(p) dp$, which is known if the distribution
$P(p)$ is known.  However, $P(p)$ is the unknown which we are going to solve
for.  In the steady state, $F_{N}(n)$ becomes identical
to the probability of the attendance in any one of the two rooms since
the two possible outcomes occur equally often on average. 
Figure 2 shows the
normalized $F_{N}(n)$ in the steady state
extracted from the numerical simulations.

In the spirit of self-consistent mean-field theories, the basic idea of the
present
formulation is to consider  the interaction between
a particular agent and the rest of the  population.
We present the formulation in a general way so that it can be readily
generalized to different variations of our model.
We consider the action of a
particular agent, say the $k$-th player, in the background of the $N-1$
other
agents.   Let $G_{N-1}^{k}(n)$ be the probability of the attendance being
$n$  in the
predicted room,  given that there are only $(N-1)$ agents participating in
the game
(i.e. excluding the
$k$-th agent).   Then $F_{N}(n)$ can be written in
terms of $G_{N-1}^{k}$ as
\begin{equation} F_{N}(n) = p_{k} G_{N-1}^k(n-1) + (1-p_{k}) G_{N-1}^k(n),
\end{equation} where $n \neq 0, N$.   Here $p_{k}$ is the $p$-value of the
$k$-th
agent at that moment. The physical meaning of Eq.(1) is transparent.   An
attendance of $n$ in room ``1" is achieved if the attendance by
the $(N-1)$ agent
background  is $n-1$ {\em and}  the $k$-th agent decides to
go
to room ``1": this leads   to the first term in Eq.(1). Alternatively the
attendance by the $(N-1)$ agent
background  is $n$
{\em
and}  the $k$-th agent decides not to go to room ``1": this leads to
the
second term in Eq.(1).

Let $\tau(p_{k})$ be the winning probability of the $k$-th agent.  Given the
probability $G_{N-1}^k(n)$, we can write
\begin{equation}
\tau(p_{k}) = p_{k} \sum_{n=0}^{(N-3)/2} G_{N-1}^k (n) + (1-p_{k})
\sum_{n=(N+1)/2}^{N-1} G_{N-1}^k (n).
\end{equation} Equation (2) says that the $k$-th agent wins if (i) the
attendance is
below
$(N-3)/2$ in room ``1" before he makes his move and he decides to go to
room
``1", thereby giving the first term or (ii)
the attendance is above $(N+1)/2$ in room ``1" before
he makes his
move and he decides not to go to room ``1", thereby giving the second term.
Since the $k$-th
agent is only characterized by his gene value $p_{k}$, $\tau(p_{k})$ can
also be
interpreted as the success rate of an agent using gene value $p_{k}$.
It follows from Eq.(1) that
\begin{eqnarray}
\sum_{n=1}^{(N-3)/2} F_{N}(n) & = & \sum_{n=1}^{(N-3)/2} \left[ p_{k}
(G_{N-1}^k(n-1) - G_{N-1}^k(n)) + G_{N-1}^k(n) \right] \nonumber \\ & = &
\sum_{n=1}^{(N-3)/2} G_{N-1}^k(n) + p_{k} G_{N-1}^k (0) -p_{k} G_{N-1}^k
(\frac{N-3}{2}). \nonumber
\end{eqnarray} Since $F_{N}(0) = (1-p_{k}) G_{N-1}^{k} (0)$,    which
follows
from
the consideration that room ``1" is empty only if the other $N-1$ agents do
not
go to room ``1" and the $k$-th agent does not go to room ``1",  we have
\begin{equation}
\sum_{n=0}^{(N-3)/2} G_{N-1}^k (n) = \sum_{n=0}^{(N-3)/2} F_{N}(n) + p_{k}
G_{N-1}^k (\frac{N-3}{2}).
\end{equation}
Similarly, we have from Eq.(1)
\begin{eqnarray}
\sum_{n=(N+1)/2}^{N-1} F_{N}(n) & = & \sum_{n=(N+1)/2}^{N-1} \left[ p_{k}
(G_{N-1}^k(n-1) - G_{N-1}^k(n)) + G_{N-1}^k(n) \right] \nonumber \\ & = &
\sum_{n= (N+1)/2}^{N-1} G_{N-1}^k(n) + p_{k} G_{N-1}^k (\frac{N-1}{2})
-p_{k}
G_{N-1}^k (N-1). \nonumber
\end{eqnarray} Since $F_{N}(N) = p_{k} G_{N-1}^{k} (N-1)$, which follows
from the
consideration that all the agents go to   room ``1" only if  all the other
$N-1$
agents go to room ``1" and the $k$-th agent goes to room ``1",  we have
\begin{equation}
\sum_{n=(N+1)/2}^{N-1} G_{N-1}^k (n) = \sum_{n=(N+1)/2}^{N} F_{N}(n) - p_{k}
G_{N-1}^k (\frac{N-1}{2}).
\end{equation}
Substituting Eqs.(3) and (4) into Eq.(2), we obtain
\begin{eqnarray}
\tau(p_{k}) & = & p_{k} \sum_{n=0}^{(N-3)/2} F_{N}(n) +  p_{k}^2 G_{N-1}^k
(\frac{N-3}{2}) \nonumber \\ & \ \ \ \ \ \ \ &
+ (1-p_{k}) \sum_{n=(N+1)/2}^N F_{N}(n) -
(1-p_{k}) p_{k} G_{N-1}^k (\frac{N-1}{2}). \nonumber
\end{eqnarray}
Using Eq.(1) to express $G_{N-1}^{k}(\frac{N-3}{2})$ in terms
of
$G_{N-1}^k(\frac{N-1}{2})$ and $F_{N}(\frac{N-1}{2})$, we then obtain
\begin{eqnarray}
\tau(p_{k}) & = & p_{k} \sum_{n=0}^{(N-3)/2} F_{N}(n) + (1-p_{k})
\sum_{n=(N+1)/2}^N F_{N}(n) \nonumber \\
&\ \ \ \ \ \ \ & + p_{k} \left( F_{N}(\frac{N-1}{2}) - 2 (1-p)
G_{N-1}^k (\frac{N-1}{2}) \right) \nonumber \\ & = &
p_{k}\sum_{n=0}^{(N-1)/2}
F_{N}(n) + (1-p_{k})\sum_{n=(N+1)/2}^{N} F_{N}(n) - 2p_{k}(1-p_{k})
G_{N-1}^k
(\frac{N-1}{2}).
\end{eqnarray}
Equation (5) separates $\tau(p_{k})$ into 3 terms, each of which
has  a physically transparent interpretation.
Consider an
``outsider", i.e. someone whose action does not affect the outcome
but
instead is only betting  on which side is the winning room according to
the
probability $p_{k}$. His winning probability is given by  the first
two
terms in Eq.(5). The third term gives the difference in the winning
probability
between an ``outsider" of the game   and an agent who actually  participates
in
the
game.  This term is negative, reflecting the fact that an agent has
a smaller probability of winning when he is actually  participating in the
game.  Consider the case in which the background population is
split evenly between room ``0" and room ``1": the $k$-th agent loses no
matter
what action he takes.
Thus the third term  represents this self-interaction term, or so-called
{\em market
impact} in financial market terminology.   The
$p_{k}(1-p_{k})$ factor means that the winning probability increases as the
gene
value $p_{k}$ deviates more from the value $1/2$,  and it produces a
symmetry
about
$p=1/2$ in $L(p)$ and $P(p)$ as shown in  Fig.1.  Note that Eq.(5) also
applies to the case when the predictor says 0:
hence it is independent of the
dynamics of the predictor which in turn is determined by the time evolution
of the
outcomes.  This further implies that the resulting $P(p)$ and $L(p)$ do not
depend on the value  of $m$ in the model.  For the present EMG,
there is a lack of an {\em a priori} perferred room: therefore the outcomes
0 and
1 will
occur similar numbers of times on the average.  In this case, the summations
in the first and second terms of Eq.(5) in the steady state
yield the value $1/2$ and  hence $\tau(p)$ becomes
\begin{equation}
\tau(p_{k}) = \frac{1}{2} - 2p_{k}(1-p_{k}) G_{N-1}^k
(\frac{N-1}{2}).
\end{equation}

In order to express the right hand side of Eq.(5)  entirely in terms of the
function $F$, we use Eq.(1) to find $G_{N-1}^k (\frac{N-1}{2})$.   From
Eq.(1),
we have
\begin{eqnarray}
%p_{k} G_{N-1}^{k}(n-1) + (1-p_{k}) G_{N-1}^k (n) = F_{N}(n) \nonumber \\
p_{k} G_{N-1}^{k}(n-2) + (1-p_{k}) G_{N-1}^k (n-1) = F_{N}(n-1).
\end{eqnarray} Subtracting the equations obtained by  multiplying Eq.(1) by
$(1-p_{k})$ and multiplying  Eq.(7) by $p_{k}$, we can eliminate $G_{N-1}^k
(n-1)$ to obtain
\begin{eqnarray} (1-p_{k}) F_{N}(n) - p_{k} F_{N}(n-1) = (1-p_{k})^2
G_{N-1}^{k}(n) - p_{k}^2 G_{N-1}^k (n-2). \nonumber
\end{eqnarray} Repeatedly applying Eq.(1), we can eliminate
$G_{N-1}^{k}(n-2)$,
$G_{N-1}^{k}(n-3)$,$\cdots$ to obtain
\begin{equation}
\sum_{j=0}^n (-1)^{n-j} F_{N}(j) (\frac{p_{k}}{1-p_{k}})^{n-j} = (1-p_{k})
G_{N-1}^k(n)\ .
\end{equation} Similarly, if we apply Eq.(1) with increasing values of $n$
instead of decreasing values of $n$, we obtain
\begin{equation}
\sum_{j=n+1}^N (-1)^{j-n-1} F_{N}(j) (\frac{1-p_{k}}{p_{k}})^{j-n-1} = p_{k}
G_{N-1}^k(n)\ .
\end{equation} Although the results are exact, in practice it makes sense to
use Eq.(8) for small
$p_{k}$ and Eq.(9) for
$p_{k} \sim 1$.
Using Eq.(8) or Eq.(9) for $n=\frac{N-1}{2}$ and substituting the result
into
Eq.(5), we obtain $\tau(p_{k})$ entirely in terms of $F_{N}(n)$, and the
label $k$
becomes irrelevant.  As mentioned, $\tau(p_{k})$ can be regarded as  the
winning
probability of an agent who is using a gene value $p$, and henceforth we
denote it by $\tau(p)$ for simplicity.

\section{Results}

In order to obtain $P(p)$ from $\tau(p)$, we note that these two quantities
are
related.  In Ref.\cite{dHR},
it was pointed  out that the stationary
distributions $P(p)$ and $L(p)$ are  proportional to each other:
\begin{equation}
\frac{P(p)}{L(p)} = constant,
\end{equation} where the right hand side is a constant independent of $p$.
Equation (10) follows from the balance between the  fluxes of agents into
and out of a region in $p$-space in the steady state.   Since an agent
using the
gene value $p$  loses $(1-2\tau(p))$ points each turn\cite{dHR},
the lifespan $L(p)$ is  given by
\begin{eqnarray}
\tau(p) = \frac{|d|}{1-2\tau(p)}. \nonumber
\end{eqnarray}
From Eq.(10), we have
\begin{equation} P(p) \propto \frac{1}{1/2 - \tau(p)},
\end{equation} with the proportionality constant determined by the
normalization
of
$P(p)$ to $\int_{0}^{1} P(p) dp = 1$.

Based on the present theory, it is straightforward to construct an iterative
calculation scheme for $P(p)$.  The steps are the
following: (a) assume a form for
$P(p)$, (b) obtain $F_{N}(n)$ by
evaluating $\overline{p}$ and the standard deviation from the
assumed $P(p)$, (c) use Eq.(5) together with Eqs.(8) and (9) to obtain
$\tau(p)$, (d) calculate $P(p)$ from $\tau(p)$ using Eq.(11) and  the
normalization condition, (e) check for convergence of $P(p)$
and, if necessary, repeat the steps until convergence is obtained.  Note
that Eq. (5)
is  employed since it is valid for  all forms of initial guess for
$P(p)$, including those which are non-symmetrical about $p=1/2$.

Results for $P(p)$ and $L(p)$ obtained by carrying out the calculation
scheme
are shown in Figs.3 and 4 together with results of  numerical simulation for
$N=51$ and $N=101$.  Note that $P(p)$, when  properly normalized, is not
sensitive to $N$, while $L(p)$ depends  on $N$.
Results from our theory are in
good agreement with numerical data.
The results for $P(p)$ as obtained in
Ref.\cite{dHR} are also shown in Fig. 3 for comparison: note that the
results of Ref.\cite{dHR}
show a plateau over a significant range of $p$ in contrast to the present
theory and the
numerical simulations.  The comparison indicates that the results from the
present theory are in
better agreement with the numerical results.  To further test the
validity of our theory, we compare results for $\tau(p)$ as a function of
$p$ with numerical data for $N=51, 101$, and $201$ in Fig. 5.
The numerical data are found by simply counting the number of times an
agent with gene value $p$ wins.
It should be noted that $\tau(p)$ provides
a better test than $P(p)$ for the validity of any theory, since many forms
of $\tau(p)$ can give rise to similar forms for $P(p)$.
In contrast to the numerical
results and those of the present theory shown in Fig. 5,
the expression for $\tau(p)$ given in Ref.\cite{dHR} gives a 
very small $\tau(p)$
for a significant range of $p$ around $p=1/2$ corresponding
to the plateau in $P(p)$.  Figure 5
suggests that the correct $\tau(p)$ in the steady state, which follows from
Eq.(5) (see also Eq.(6)), has the form
$\tau(p) \sim 1/2 - {\cal A}(N) p (1-p)$
where ${\cal A}(N)$ is an $N$-dependent constant which decreases with $N$
as $1/\sqrt{N}$. Such a scaling with $N$ makes sense from random walk
arguments.

\section{Discussion}

We have presented a theory of the EMG based on the consideration
of a particular agent in the environment formed by the rest of the
population.  The winning probability $\tau(p)$ is given in
terms of the population distribution in one of the rooms.  By relating
the population distribution, the winning probability and the lifespan,
an iteration scheme is set up for calculating the frequency
distribution of gene values $P(p)$.  Results for $P(p)$, $L(p)$ and
$\tau(p)$ are in good agreement with numerical data.

The present formalism can be used to describe different versions
of the EMG.
For example, a generalization of the EMG 
was recently introduced where
the winning
`room' (i.e. winning decision) was assigned according to whether the
attendance was lower than a
certain cutoff
\cite{freezing}. For this case,
one can modify the limits in the summations in Eq.(2) and
carry out the calculations accordingly.  We emphasize that Eq.(5) is
applicable
even if the steady state $P(p)$ is not symmetric about $p=1/2$.
An interesting feature in this generalized EMG model is that when
the cutoff percentage deviates significantly from $1/2$ and
becomes smaller (or larger) than a critical value, the steady state $P(p)$
takes on a form which depends on the initial distribution of $p$.
In particular, the population distribution $P(p)$ freezes - no further
modification of gene values arises as time evolves for large (or
small) enough value of the cutoff. This phenomenon is discussed in more
detail in  Ref.\cite{freezing}.  Another generalization
is to modify the way in
which the $p$-value is updated\cite{ceva2}. Future work will focus on
application of the
present theoretical approach to such generalizations of the simple minority
game set-up.

\begin{figure}
\caption{The lifespan $L(p)$, which is the average duration between
modifications for a gene value $p$, as a function of gene value $p$ for
$m=1,2,\cdots,8$.  The inset shows the distribution of gene values $P(p)$ as
a function of $p$ for different values of $m$.  Both $L(p)$ and
$P(p)$ are insensitive to $m$. The other parameters are $N=101$, $d=-4$ and
$R=0.2$.}
\vspace*{0.3 true in}

\caption{The probability of the attendance in one of the two
rooms in the steady state, which is identical to
$F_{N}(n)$, obtained by numerical simulations.  The parameters are $N=101$,
$m=3$, $d=-4$ and $R=0.2$.  It is approximately a gaussian distribution as
expected
from the central limit theorem.}
\vspace*{0.3 true in}

\caption{The frequency distribution of the gene values $p$ as a function
of $p$ for $N=101$ and $N=51$ (inset).  The other parameters are $d=-4$
and $R=0.2$.  The dotted lines are the data from numerical simulation.
The solid lines give the results of the present theory.  The dashed
lines give the results of the theory proposed in Ref.[12].}
\vspace*{0.3 true in}

\caption{The lifespan $L(p)$ as a function of $p$ for $N=101$ and
$N=51$ (inset).  The dotted lines are the data from numerical simulation.
The solid lines give the results of the present theory.  Other parameters
are the same as those in Fig.3.}
\vspace*{0.3 true in}

\caption{The winning probability $\tau(p)$ as a function of $p$ for
different values of $N$.  The solid lines give the results of the
present theory while the dotted and dashed
lines are results from numerical simulations.
The three sets of lines from top to bottom at $p=1/2$ correspond to
$N=201$, $101$, and $51$, respectively.  Other parameters are the same as
those
in Fig.3.}
\end{figure}

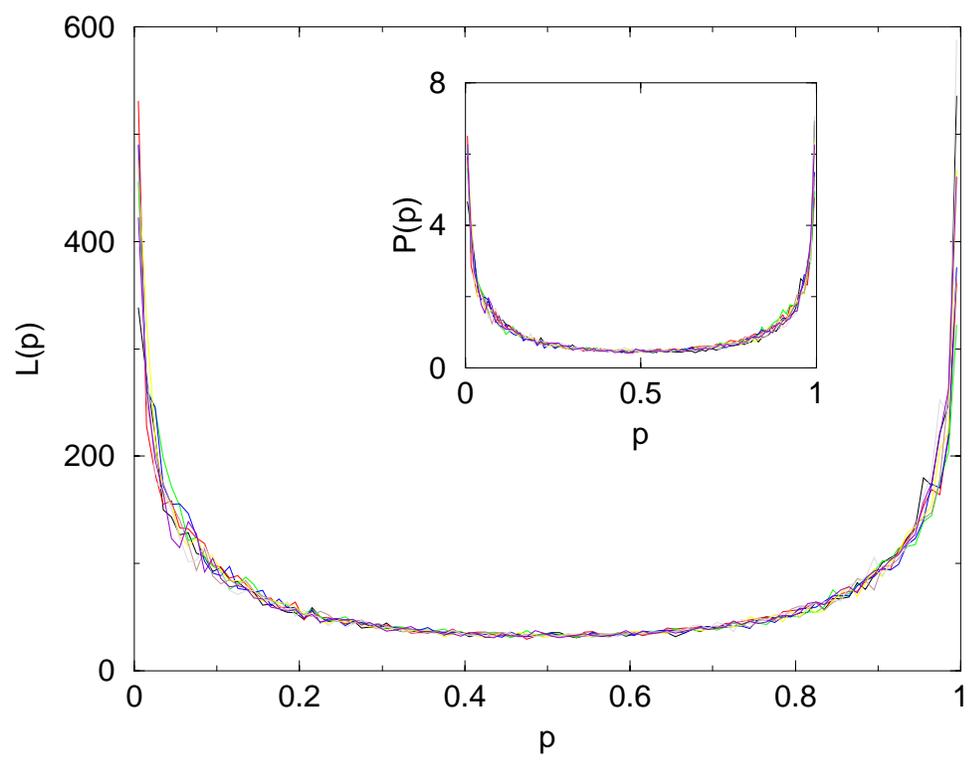
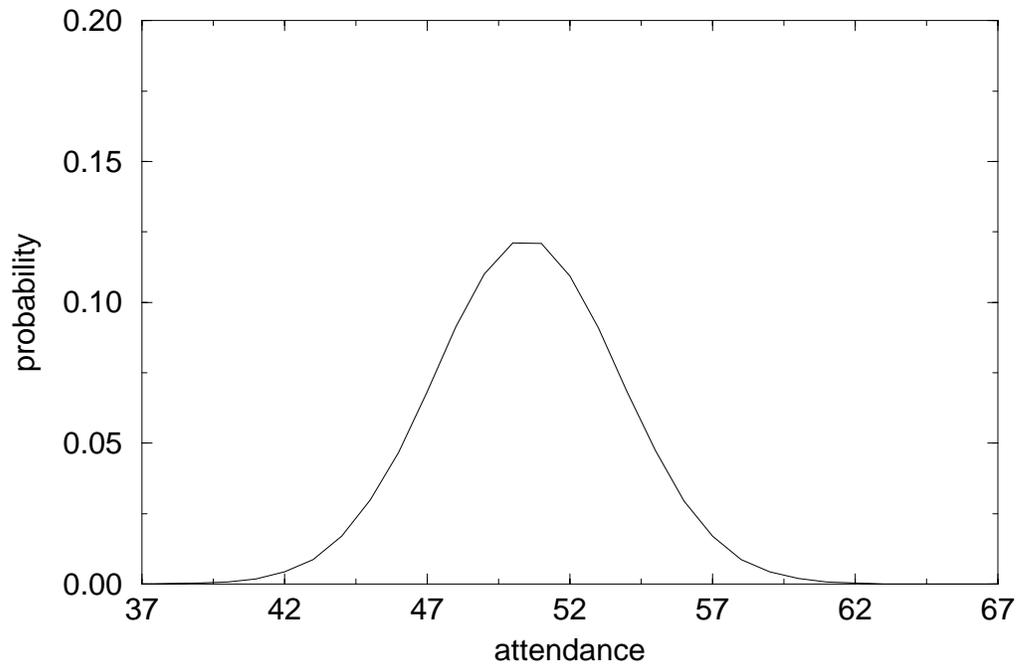
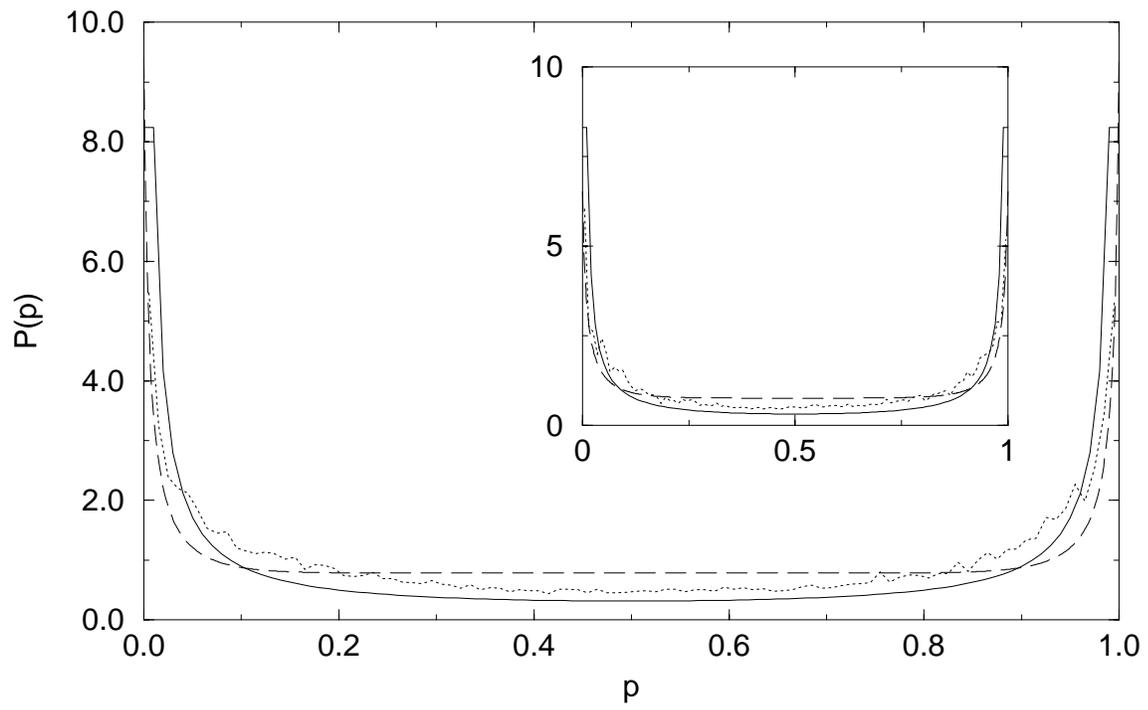
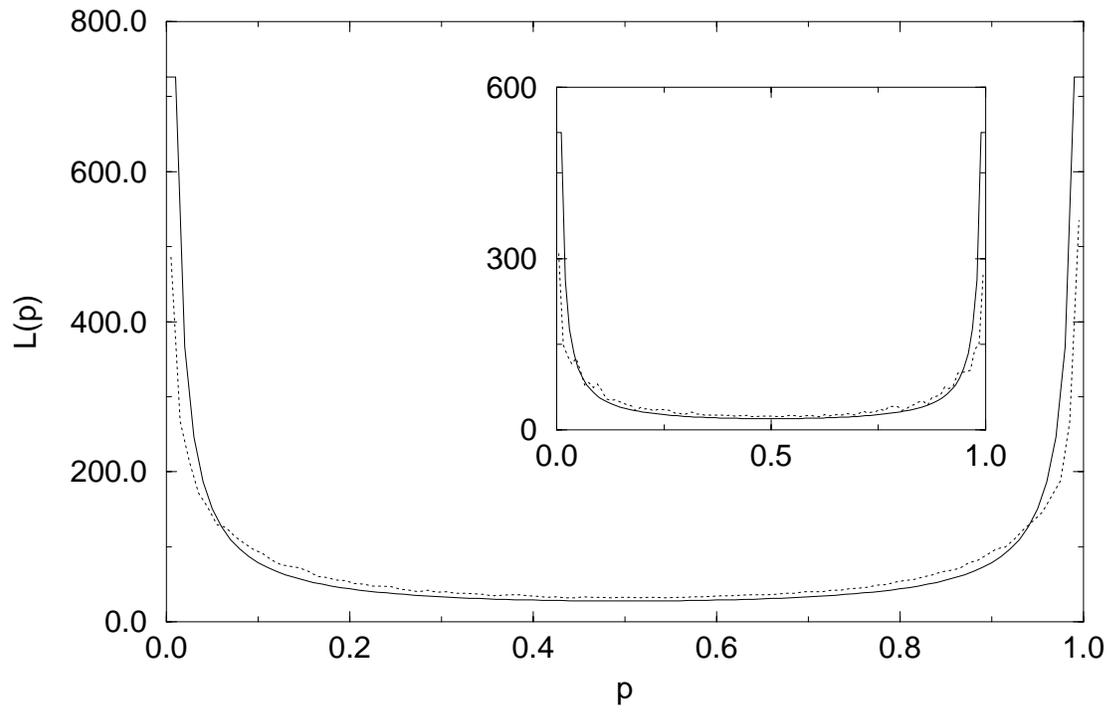
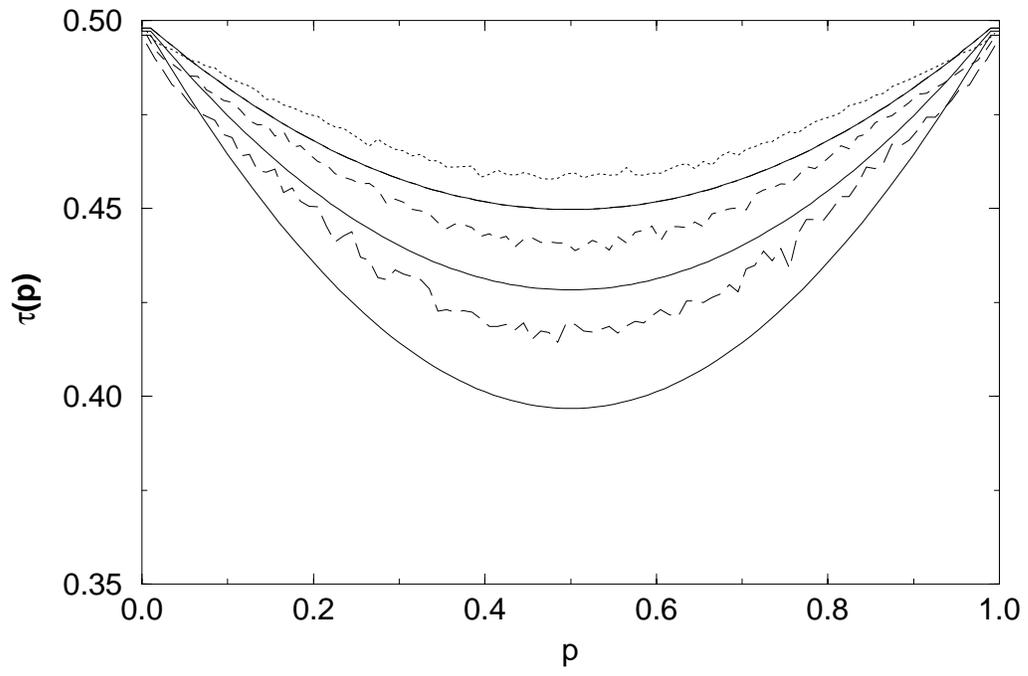
\end{document}